\documentclass[aps,prl,amsmath,amssymb,reprint, twocolumn,superscriptaddress, reprintnumbers,showpacs,intlimits,longbibliography]{revtex4-1}
\pdfoutput=1
\bibpunct{[}{]}{,}{n}{}{}

\usepackage[utf8]{inputenc}
\usepackage{bm,latexsym,mathrsfs,enumerate}
\usepackage[dvipsnames]{xcolor}
\usepackage{mathtools}
\usepackage{makecell}
\usepackage{multirow}
\usepackage[breaklinks=true,unicode=true,urlcolor = blue,colorlinks = true,citecolor = blue,linkcolor = blue]{hyperref}
\usepackage{graphicx}
\usepackage{ragged2e}
\usepackage[export]{adjustbox}
\usepackage[final]{pdfpages}
\usepackage{tikz}
\makeatletter
\AtBeginDocument{\let\LS@rot\@undefined}
\makeatother
\begin{document}

\title{Nonlinear Decay of Quantum Confined Magnons in Itinerant Ferromagnets}

\author{ Kh.~Zakeri}
\email{khalil.zakeri@partner.kit.edu}
\affiliation {Heisenberg Spin-dynamics Group, Physikalisches Institut, Karlsruhe Institute of Technology, Wolfgang-Gaede-Str. 1, D-76131 Karlsruhe, Germany}
\author{A. Hjelt}
\affiliation {Heisenberg Spin-dynamics Group, Physikalisches Institut, Karlsruhe Institute of Technology, Wolfgang-Gaede-Str. 1, D-76131 Karlsruhe, Germany}
\author{I. V. Maznichenko}
\affiliation {Department of Engineering and Computer Sciences, Hamburg University of Applied Sciences, Berliner Tor 7, D-20099 Hamburg, Germany}
\author{P. Buczek}
\affiliation {Department of Engineering and Computer Sciences, Hamburg University of Applied Sciences, Berliner Tor 7, D-20099 Hamburg, Germany}
\author{A.~Ernst}
\affiliation {Institute for Theoretical Physics, Johannes Kepler University, Altenberger Str. 69, A-4040 Linz, Austria}
\affiliation{Max-Planck-Institut f\"ur Mikrostrukturphysik, Weinberg 2, D-06120 Halle, Germany}
\begin{abstract}
Quantum confinement leads to the emergence of several magnon modes in ultrathin layered magnetic structures. We probe the lifetime of these quantum confined modes in a model system composed of three atomic layers of Co grown on different surfaces. We demonstrate that the quantum confined magnons exhibit nonlinear decay rates, which strongly depend on the mode number, in sharp contrast to what is assumed in the classical dynamics. Combining the experimental results with those of linear-response density functional calculations we provide a quantitative explanation for this nonlinear damping effect.  The results provide new insights into the decay mechanism of spin excitations in ultrathin films and multilayers and pave the way for tuning the dynamical properties of such structures.
\end{abstract}

\maketitle

Understanding the processes behind the excitation and relaxation of spin excitations in low-dimensional magnetic structures is one of the most intriguing research directions in solid-state physics. A detailed knowledge on the fundamental mechanisms involved is such processes is the key to understand many different phenomena. Examples are ultrafast magnetization reversal by ultrashort magnetic field pulses \cite{Tudosa2004} or by torque transfer from spin-polarized currents \cite{Slonczewski1996, Katine2000, Garello2014}, vortex core gyration driven magnon emission \cite{Wintz2016}, subpicosecond demagnetization by ultrafast photon pulses \cite{Beaurepaire1996, Stanciu2007, Koopmans2009, Lambert2014}, and strong spin-dependence of image potential states at ferromagnetic surfaces \cite{Schmidt2010}. In addition to its fundamental impact, a complete understanding of magnetic relaxation mechanisms is of great importance for designing efficient spin-based devices as the power consumption of such devices is determined by the magnetic damping  \cite{Zakeri2014,Walowski2016, Zakeri2018a}.

The excited state of a magnetic system is described by magnons, the quanta of spin waves. The relaxation of such an excited state can, in principle, involve the dissipation of magnetic energy in several different ways. In the most common approach, based on the classical dynamics, damping is described by a phenomenological damping parameter, commonly referred to as Gilbert damping \cite{Sparks1970, Gilbert2004, McMichael2004}.
This description is only valid in the case of uniform ferromagnetic resonance, i.e. the magnons with zero wavevector, under some circumstances \cite{McMichael2003, Lindner2003, Heinrich2003, Woltersdorf2004, Zakeri2007}. For magnons with a nonzero wavevector it has been assumed that the relaxation of a magnon with a certain wavevector can involve its dissipation to other magnons with different wavevectors (multi-magnon scattering process) or,  in the case of itinerant magnets, their dissipation into the single-particle electron-hole pair excitations, known as Stoner excitations. In both cases one would expect an increase of the magnon decay rate with the magnon energy, since usually the density of both magnon as well as Stoner states increases with energy.

In structurally well-defined low-dimensional magnetic structures one would expect additional magnon modes as a result of quantum confinement.  For example in ultrathin magnetic films or multilayers composed of a finite number of atomic layers $N$ one can show that there exist $n=0, 1, ...., N-1$ different magnon modes, as a result of quantum confinement in the direction perpendicular to the structure \cite{Balashov2006,Gao2008, Rajeswari2014, Ibach2014, Zakeri2013, Balashov2014, Buczek2018}. These magnon modes spread in the energy--momentum space \cite{Taroni2011,Bergqvist2013,Eriksson2016}. The dispersion relation of  all confined magnon modes has recently been probed \cite{Chen2017, Zakeri2020, Zakeri2021}. However, the relaxation mechanism of these magnon modes remains hitherto unexplored.

 In this Letter we report on the decay rate of the quantum confined magnon modes in a model system. We show that the quantum confined magnons exhibit  decay rates, which are nonlinear in energy, and strongly depend on the magnons' mode number $n$. This observation is in contrast to what is commonly discussed in the framework of the classical dynamics. Combining the results of linear-response time-dependent density functional theory calculations with adiabatic spin dynamics calculations we provide a quantitative explanation for the damping of quantum confined magnons.

\begin{figure*}[t!]
	\centering
	\includegraphics[width=1.5\columnwidth]{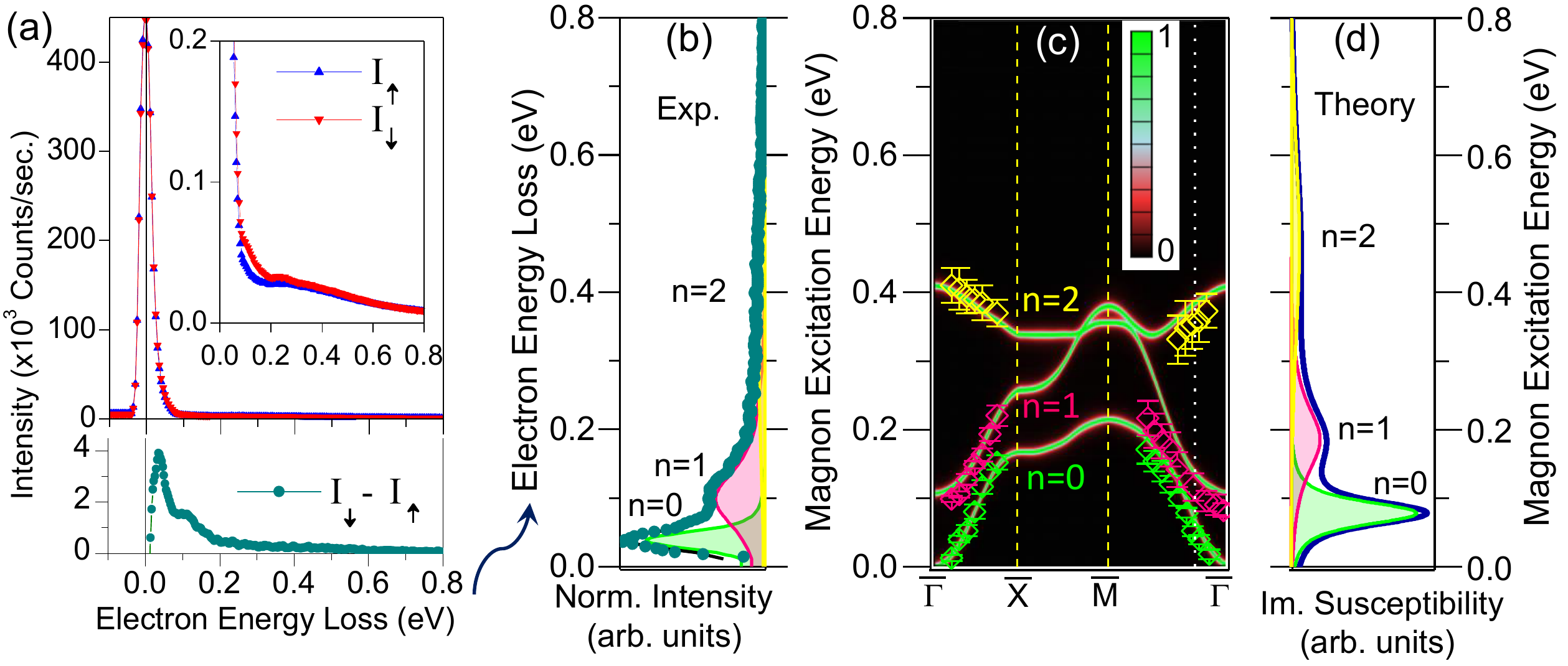}
	\caption{(a) SPEELS spectra recorded on 3 ML Co/Ir(001) at ($q_x$, $q_y$)=(0.3 \AA$^{-1}$, 0.3\AA$^{-1}$). $I_{\downarrow}$ ($I_{\uparrow}$) represents the spectrum when the spin polarization of the incoming electron beam is parallel (antiparallel) to the magnetization. The difference spectrum $I_{\downarrow}-I_{\uparrow}$ is shown in the lower panel. (b) The difference spectrum and the fits used to extract the dispersion relation and the lifetime of different quantum confined magnons indicated by $n=0$, $1$ and $2$.  (c) The dispersion relation of all confined magnon modes. The experimental data are shown by symbols and the calculated magnon Bloch spectral function using our adiabatic approach is presented as the color map. The dotted line shows the place in the surface Brillouin zone, where the spectra shown in (a) and (b) are recorded. (d) The imaginary part of the dynamical susceptibility calculated at (0.3, 0.3) by LRTDDFT.}
	\label{Fig:Fig1}
\end{figure*}

Experiments are performed by means of spin-polarized high-resolution electron energy-loss spectroscopy (SPEELS) on ultrathin Co films with a thickness of three atomic layers epitaxially grown of Ir(001), Ir(111) and Cu(001). Figure \ref{Fig:Fig1}(a) shows typical SPEELS spectra recorded on a 3 monolayer (ML) Co film on Ir(001). The spectra are recorded for the two possible incoming spin states. $I_{\downarrow}$  ($I_{\uparrow}$)  represents the intensity spectrum when the spin polarization of the incoming beam is parallel (antiparallel) to the sample magnetization. The difference spectrum, shown in the lower panel, contains all the possible spin-flip excitations of down to up, including magnons \cite{Prokop2009,Zakeri2013}. The data are recorded at ($q_x$, $q_y$)=(0.3 \AA$^{-1}$ , 0.3 \AA$^{-1}$), corresponding to $q_{\parallel}=0.42$   \AA$^{-1}$ along the $\bar{\Gamma}$--$\bar{\rm{M}}$ direction. In order to unambiguously determine the magnon excitation energy and the lifetime, the difference spectra recorded at different wavevectors are fitted with three lines, corresponding to the expected three magnon modes of the system. Each line includes a Lorentzian lineshape, convoluted with the experimental broadening \cite{Zhang2012, Zakeri2012}. An example is shown in Fig. \ref{Fig:Fig1}(b), where the experimental difference spectrum ($I_{\downarrow}-I_{\uparrow}$) is shown together with the fits. The magnon dispersion relation was measured by probing the magnons at different wavevectors and along different symmetry directions of the surface Brillouin zone and the results are summarized in Fig. \ref{Fig:Fig1}(c). For a quantitative description the magnon properties were calculated based on first-principles. It has been shown that a quantitative description of the experimental magnon bands can only be provided when spin-dependent many-body correlation effects on the majority Co spins are taken into consideration \cite{Chen2017, Qin2019}.
In Fig. \ref{Fig:Fig1}(c) the magnonic band structure calculated using this approach is presented as the color map. The band structure is presented by plotting the magnon Bloch spectral function. Since the adiabatic approach does not account for the decay of magnons, the spectral function exhibits sharp peak at the places where different magnon modes exist. The approach accounts, however, for all the details of the geometrical structure (e.g., the reconstruction of the Ir surface) and provides an unambiguous way for the determination of magnon properties e.g., their real space localization and density of magnon states \cite{Zakeri2021}.

In order to understand the decay mechanism of magnons we calculated, based on first principles, the frequency $\omega$ and momentum $\mathbf{q}$ dependent transverse dynamical spin susceptibility $\chi (\omega, \mathbf{q})$, using linear-response time dependent density functional theory (LRTDDFT) \cite{Mook1979, Savrasov1998,Buczek2011a,Buczek2011b}. In Fig. \ref{Fig:Fig1}(d) we provide an example of $\mathfrak{Im}\chi (\omega, \mathbf{q})$  at ($q_x$, $q_y$)=(0.3, 0.3). Since in this approach both magnons and Stoner excitations are taken into account, this quantity can directly be compared to the difference spectrum shown in Fig. \ref{Fig:Fig1}(b) \cite{Callaway1981,Hong2000,Muniz2003,Costa2006,Muniz2008,Costa2008,Costa2010a}. In order to have a better comparison, we convolute the results with the experimental resolution. Similar to the experiment, one observes all three confined magnon modes of the system and different magnon modes exhibit different decay rates.

\begin{figure}[t!]
	\centering
	\includegraphics[width=0.999\columnwidth]{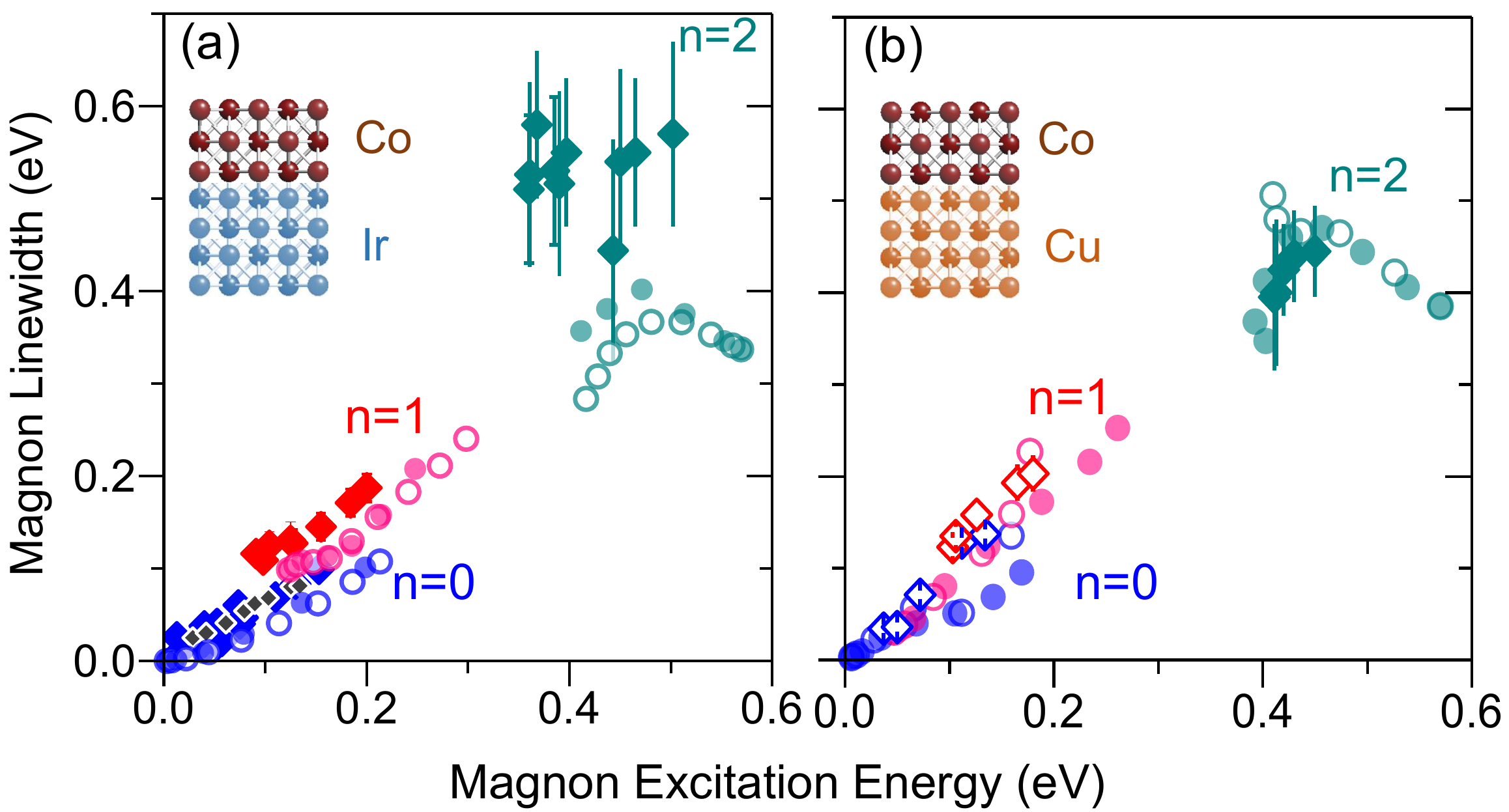}
	\caption{The magnon linewidth versus energy for all quantum confined magnon modes in 3 ML Co on (a) Ir(001) and (b) Cu(001). The experimental results are shown by diamonds and the results of LRTDDFT calculations are shown by circles. The filled (open) symbols represent the data along the $\bar{\Gamma}$--$\bar{\rm M}$ ($\bar{\Gamma}$--$\bar{\rm X}$) direction. The black diamonds in (a) represent the results of the $n=0$ magnon mode of 3 ML Co/Ir(111).}
	\label{Fig2}
\end{figure}

In order to quantify the decay rates of different quantum confined magnon modes we carefully analyze the intrinsic broadening of the modes and the results are summarized in Fig. \ref{Fig2}(a), indicating that different quantum confined magnon modes show different decay rates.
Moreover, to address the effects of the symmetry and hybridization effects between the electronic structures of the film with those of the underlying substrate similar experiments were performed on Co films grown on the Cu(001) and Ir(111) surfaces \cite{Zakeri2013,Meng2014,Chuang2014,Zakeri2017}. The results are presented along with those of the Co on Ir(001) system. We note that due to the geometrical consideration these systems possess different magnonic band structures. The main aim here is to understand the decay rate of different magnon modes, with respect to the mode number and mode energy. In a similar manner we also analyze the line broadening of different magnon modes as calculated by LRTDDFT and the results are summarized in Fig. \ref{Fig2}, indicating a very good agreement with the experimental results.
A careful analysis of the experimental and theoretical results indicates that, (i) different magnon modes possess different decay rates, (ii) for all studied systems the magnon mode with $n=0$ exhibits the lowest damping among all the others, and (iii) the magnon linewidth versus energy is nonlinear, which becomes more pronounced for the higher order modes.

Generally the main source of damping in itinerant ferromagnets, such as systems studies here, is the result of the decay of these collective modes into the single-particle Stoner excitations, a mechanism known as Landau damping. It has been shown that in the case of ultrathin magnetic films on nonmagnetic substrates, the hybridizations of the electronic states of the film with those of the substrate can open additional decay channels and lead to  stronger damping of  the magnons, meaning that the Landau damping in layered ferromagnets on nonmagnetic substrates can be very complex, compared to that of the single element bulk ferromagnes \cite{Buczek2011a,Buczek2011b, Qin2015, Qin2017}. Hence, in order to adequately describe the magnon damping a proper description of the electronic structures is required. In our LRTDDFT calculations, we start with the electronic structures calculated based on the experimental geometrical structure. Since the approach treats both magnons and Stoner excitations on the same basis, the Landau damping is fully taken into consideration.

\begin{figure}[t!]
	\centering
	\includegraphics[width=0.95\columnwidth]{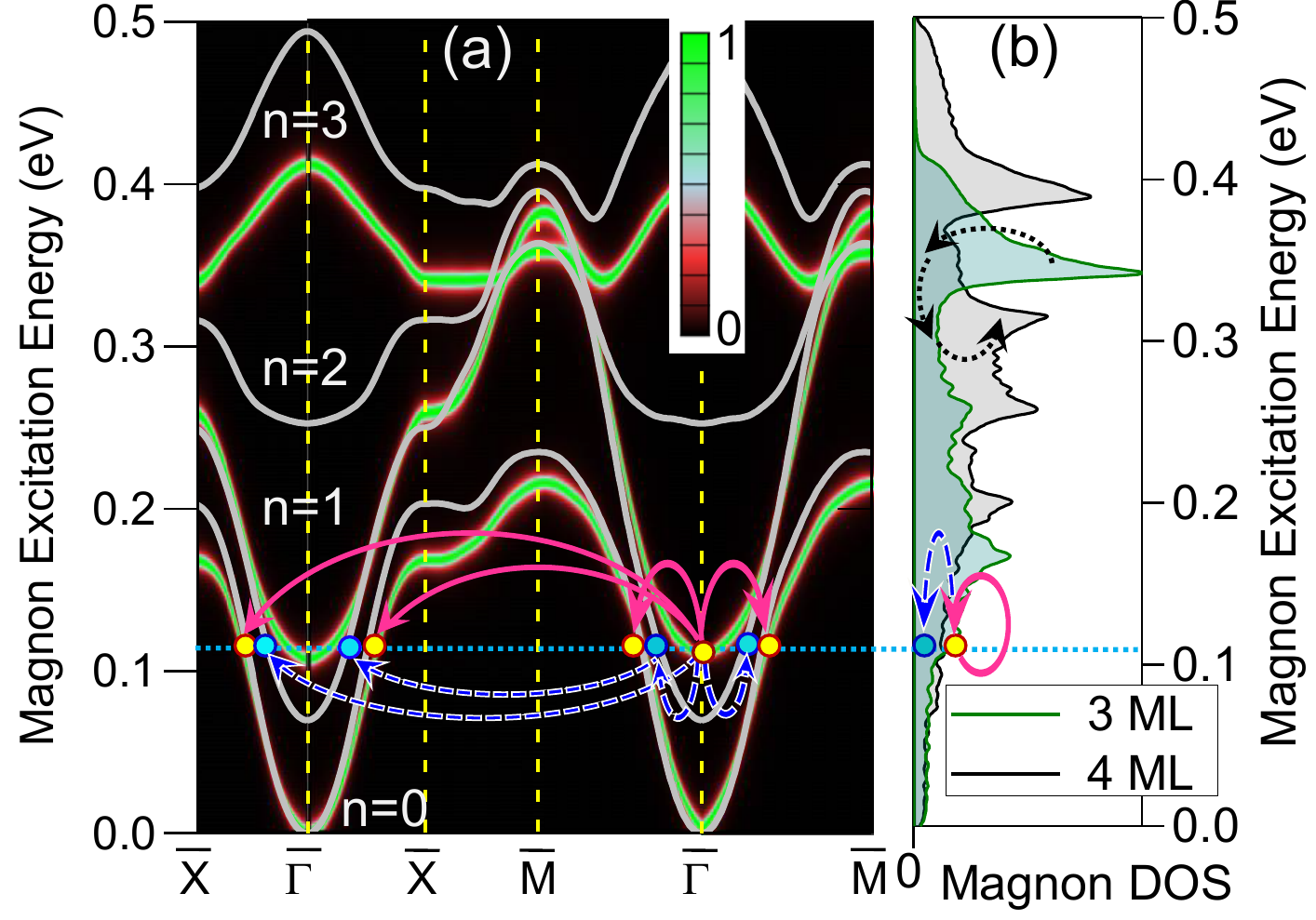}
	\caption{(a) The magnon Bloch spectral function of 3 ML Co on Ir(001) based on adiabatic calculations. The magnonics bands of a 4 ML film are also shown by the light-gray curves. The multimagnon scattering process of the $n=1$ magnon mode within the 3 ML terraces are shown by the solid arrows. The decay process within the terraces of different thicknesses (3 and 4 ML) are shown by the dashed arrows. (b) The magnonic density of states for a 3 ML and 4 ML film calculated based on the adiabatic approach. The direct decay of the $n=1$ magnon mode of the 3 ML film to the $n=0$ and $n=1$ of a 4 ML film is illustrated with the dashed arrows. The indirect decay of the $n=2$ magnon mode of the 3 ML film to the $n=2$ and $n=1$ modes of 4 ML film is schematically shown by the dotted black arrows.}
	\label{Fig3}
\end{figure}

Looking at the results shown in Fig. \ref{Fig2}, for each magnon mode the linewidth increases with energy in a nonlinear fashion. The increase of the magnon linewidth with energy is due to the fact that the probability that a magnon decays into a Stoner pair increases with energy. This probability depends on the details of the involved electronic bands and hence is not a simple linear function with energy. This scenario is even more complex when considering the fact that the electronic states of the film hybridize very strongly with those of the substrate. The so-called Landau hot spots, where the decay of magnons to Stoner excitations takes place become increasingly important for higher energy magnons  \cite{Buczek2011a,Buczek2011b, Qin2015}. As it is apparent from Fig. \ref{Fig2}  both the experimental and theoretical results indicate that different magnon modes exhibit different (nonlinear) decay rates. This implies that the origin of the mode-dependent decay rate lies in the Landau damping. The effect can be understood based on the fact that for the $n \neq 0$ magnon modes the perpendicular magnon momentum is nonzero ($q_{\perp}\neq0$). Hence the Stoner pairs with nonzero perpendicular momentum will enter the picture and, therefore, the possibility of the decay into such modes shortens the lifetime of magnons with $n \neq 0$.
In the calculations based on LRTDDFT both the magnons and Stoner excitations are put on an equal footing and should therefore adequately describe the decay rates of different magnon modes caused by their dissipation into the Stoner excitations. Hence this effect is clearly observed in the calculated decay rates.

The second possible decay channel is the decay of a certain magnon mode to the other possible magnon (or phonon) modes which share the same energy. As an example in Fig. \ref{Fig3}(a) the processes of magnon decay of the $n=1$ magnon mode at the $\bar{\Gamma}$-point is shown by the solid arrows. In addition, small variations in the film thickness can lead additional magnon modes, which may also share the same energy with this mode \cite{Michel2016}. In order to mimic this effect in Fig. \ref{Fig3}(a) we also show the magnonic bands of a film with the same geometrical structure but with a thickness of 4 ML. The results are shown by the light-gray color in Fig. \ref{Fig3}(a). Since the $n=1$ mode can also degenerate with the modes of such a system one needs to consider such decay rates. The decay rates of this kind are shown by the dashed arrows in Fig. \ref{Fig3}(a). In the LRTDDFT calculations such effects are not taken into account.
In order to generalize the decay process of a certain magnon mode to all the other possible quasiparticles the damping may be written as \cite{Qin2015, Park2020}

\begin{eqnarray}
\Gamma(\mathbf{q}, \omega) &=&\sum_{n,m} \int_{\Omega_{BZ}} \mathcal{P}_{\mathbf{k}}^{n}\mathcal{P}_{\mathbf{k}-\mathbf{q}}^{m}\mathbf{\delta}\left(E-E_{\mathbf{k}}^{n}-E_{\mathbf{k}-\mathbf{q}}^{m}\right)d\mathbf{k},
\label{Eq1}
\end{eqnarray}
where $\mathcal{P}_{\mathbf{k}}^{n}$  and $\mathcal{P}_{\mathbf{k}-\mathbf{q}}^{m}$ denote the probability of finding a magnon, phonon or electron with the wavevector $\mathbf{k}$ in the $n$th and $\mathbf{k}-\mathbf{q}$ in the $m$th band, $E_{\mathbf{k}}^{n}$ ($E_{\mathbf{k}-\mathbf{q}}^{m}$) describes the energy dispersion of the $n$th ($m$th) quasiparticle band. The term shall account for all the possible decay channels, which a magnon with the energy $\hbar \omega$ and momentum $\mathbf{q}$ can decay
into single-particle Stoner pairs, other magnons and phonons, satisfying
the energy conservation rule $E_{\mathbf{q}}=\hbar \omega=E_{\mathbf{k}}-E_{\mathbf{k}-\mathbf{q}}$.  Note that in the case of electronic bands the transition should also account for the conservation of magnon's total angular momentum, meaning that only the transitions between the bands with opposite spins should be considered.
In order to describe the decay of magnons into bosonic quasiparticles, e.g., phonons and other possible magnon states, according to Eq. (\ref{Eq1}), one should be able to analyze the different possibilities of such decays in the energy--momentum space. In metallic ferromagnets the phonon energies are rather low ($<$ 20 meV) and the magnon-phonon coupling is rather weak. Hence, the magnon decay by phonons becomes of minor importance for high-energy magnons. The main decay channel of magnon-boson kind is the one associated with their decay into other magnon modes. In order to estimate the strength of such decay rates we calculated the magnonic density of states (DOS). The results of such calculations are presented in Fig. \ref{Fig3}(b). We also present the magnon DOS of a film composed of four atomic layers. The probability of finding magnons in a given state (indicated by circles) can therefore be simply estimated by analyzing the magnon DOS. The magnon--magnon decay is directly proportional to the number of initial and final magnon states, which may contribute to such a process ($\mathcal{P}_{\mathbf{k}}^{n}$  and $\mathcal{P}_{\mathbf{k}-\mathbf{q}}^{m}$ ). If such states are largely available the magnon decay can occur with a large probability. Looking at the data presented in Fig. \ref{Fig3} one realizes that such decay process can occur with a large probability, since there are enough initial and final magnon states, which can contribute to this kind of magnon decays (solids arrows in Fig. \ref{Fig3}). In addition to the intrinsic magnon-magnon decay, the variation in the film thickness can also lead to a magnon decay. For example if the film is composed of terraces with the thickness of 3 and 4 ML, the $n=1$ magnon mode of the 3 ML terraces can decay into the $n=1$ and $n=0$ magnon modes of the terraces with the thickness of 4 ML. Such a process can happen with a large probability as shown by the dashed arrows in Fig. \ref{Fig3}. Interestingly the $n=2$  mode of the 3 ML region can, in principle, decay to the other modes of the 4 ML region via an inelastic process in which the magnons are decayed in a two step process [dotted arrows in Fig. \ref{Fig3}(b)]. The process can occur with a high probability because near the states of the $n=2$  mode of the 3 ML region there exist a large number of states caused by the $n=2$ and $n=1$ modes of the 4 ML region. Due to the reconstruction of the Ir(001) surface the roughness of Co films on this surface is larger than that of the Co films on Cu(001). This leads to a larger magnon decay of this kind and explains the larger experimental linewidth of this mode as compared to the Co/Cu(001) system and also to the results of LRTDDFT.

In summary, aiming on a fundamental understanding of the decay processes of quantum confined magnons in layered ferromagnets, we investigated the lifetime of these excitations in a model system composed of 3 ML Co grown on different surfaces over a wide rage of energy and momentum. It was observed that the quantum confined magnons exhibit nonlinear decay rates. The decay rates strongly depend on the mode number. In phenomenological approach of classical dynamics the decay rate is assumed to be linear. Such an assumption is not valid for the quantum confined magnons. Combining the experimental results with those of LRTDDFT calculations we provide a quantitative explanation for this nonlinear damping.  In addition, since the quantum confinement leads to the emergence of several magnon branches, the decay processes as a result of magnon-magnon scattering become also important. These multimagnon decay processes become stronger due to variations in the film thickness. Our results indicate that the main source of damping in layered structures made of itinerant ferromagnets is due to the Landau damping as a result of their decay into Stoner excitations. Hence in order to design layered ferromagnets with low damping, first the electronic structure should be tuned such that the Landau damping is suppressed. Moreover, atomically flat films are required to achieve a low damping. In addition to the fact that our results provide new insights into the decay mechanism of spin excitations in ultrathin films and multilayers, they provide guideline regarding how the dynamical properties of layered structures can be tuned.

\section*{Acknowledgements}
Kh.Z. acknowledges funding from the Deutsche Forschungsgemeinschaft (DFG) through the Heisenberg Programme ZA 902/3-1 and ZA 902/6-1 and the DFG Grant ZA 902/4-1.  Kh.Z. thanks the Physikalisches Institut for hosting the group and providing the necessary infrastructure.

%

\end{document}